\documentstyle[aps,preprint]{revtex}
\tightenlines

\begin{document}
\title{Skyrmion dynamics in quantum Hall ferromagnets}
\author{A.Villares Ferrer and A. O. Caldeira\thanks{E-mail: caldeira@ifi.unicamp.br}}
\address{Instituto de F\'{\i}sica ``Gleb Wataghin'',\\
Departamento de F\'{\i}sica do Estado S\'{o}lido e Ci\^{e}ncia dos Materiais,
\\
Universidade Estadual de Campinas, 13083-970, Campinas, SP, Brasil.}
\date{\today}
\maketitle

\begin{abstract}
Exploring a classical solution of the non-linear sigma model for a quantum Hall ferromagnet, 
a skyrmion-magnon effective hamiltonian is obtained via the collective coordinates method. 
Using the Feynman-Vernon functional integral formalism for this model we find the   
temperature dependent transport coefficients which characterize a single skyrmion
dynamics. 
\end{abstract}

\section{Introduction}

The concept of skyrmion was introduced by T. Skyrme when he represented 
the nucleon, a fermion, as a topological excitation (soliton) of the bosonic pion field 
\cite{skyrme}. In the quantum Hall ferromagnet, skyrmions are the solitons in the 
spin space of the non linear field theory for electrons, and have spin much 
larger than $\frac{1}{2}$.\\

The theory for the skyrmions, in the 2-D electron gas subject to strong magnetic
fields, was developed by Shondi et al \cite{shondi} based on an effective generalized 
non linear sigma model ($O(3)$) which was deduced from the Landau-Ginzburg theory of spinful 
electrons \cite{lee}. 
Shondi's conjecture was experimentally confirmed by  
Barrett and co-workers in the optically pumped nuclear magnetic resonance 
experiments \cite{barrett}.
From both works, \cite{shondi} and \cite{barrett}, it can be seen that
this topological excitations are detectable in the quantum Hall ferromagnet when the 
exchange interaction dominates over the Zeeman coupling. 
This situation is possible due to the smallness of the effective Land\'e factor ($g$) 
in some particular materials, as in the $GaAs$ case. 
If this is so, the simple creation or destruction of an electron in the 
fully occupied spin polarized Landau level, $\nu=1$ for instance, do not produce the 
low energy configuration to be considered the lowest lying charged excitation. 
On the contrary, a non-uniform cylindrically symmetric spin configuration is produced, 
in which the spin at the centre is pointing in the wrong direction and smootly rotates as 
one moves radially outwards until it reaches the right direction according to the polarized 
state of the 2-dimensional electron gas. Later on, Hartree-Fock numerical
calculations \cite{fertig} showed that the energy of the quantum Hall skyrmion excitation 
is always smaller than the excitation energy of the localized spin $\frac{1}{2}$ 
quasiparticle and quasihole, supporting the idea that the skyrmions are for 
sure the relevant quasiparticles in the $g \rightarrow 0$ limit.\\

In practice it is convenient to regard the skyrmion as a topological object. From 
this point of view they can be described as the non-singular mapping of the compactified 
2-D real space ($S_{2}^{phy}$) onto the 2-D internal space ($S_{2}^{int}$) whose elements 
are the unitarily normalized spin vectors which cover a 2-D spherical surface.
All non-singular maps of $S_{2}$ onto $S_{2}$ can be classified into homotopy
classes that form a group isomorphic to the group of integers. And, it is using 
this set of integers, that we define the topological charge or Pontryaguin index $Q$, 
that classifies the skyrmions. This topological charge can be shown to be
a simple measure of the degree of mapping of one surface onto the other and, as a 
conserved number, helps in finding explicit expressions for the spin configurations
(see \cite{poly}, for instance).\\

Until now many theoretical studies were devoted to the properties of the quantum Hall 
skyrmions. Most of them are done following two main approches; the quantum 
computational calculations and the non-linear semi-classical field theory. 
In the first case, the quantum Hartree-Fock calculations are shown to be in excellent 
agreement with exact diagonalizations for small skyrmion sizes, as is demostrated 
in \cite{critical}. 
However, when hydrostatic pressure is applied to the quantum Hall ferromagnet
the $g$ factor almost vanishes. In that case there is strong experimental evidence 
\cite{larg} that skyrmions of large size (about 33 reversed spins) are dominant. 
In this particular situation computational calculations are not appropriate, 
making the development of new approaches in the semi-classical field theoretical methods 
the natural way out to explain the experimental data.\\

In this paper, starting from a specific Lagragian density \cite{shondi}, we investigate 
the influence of the ferromagnetic spin wave modes  in the temperature dependent dynamical 
properties of an isolated skyrmion. Our model basically consists of an effective field 
theory in which the single skyrmion momentum is coupled to the magnon bath total momentum. 
To begin with, sec II is devoted to obtaining the ferromagnetic spin wave modes by computing 
the quantum fluctuations of the quantum Hall ferromagnet around the non-linear sigma model 
classical solutions. Later on, using the standard collective coordinate method a second 
quantized  hamiltonian that simulates the scattering of the spin waves by the skyrmion in 
$2$-D is obtained.\\

In sec. III, using the well known Feynman-Vernon formalism we obtain an effective 
equation of motion for the centre of mass of the skyrmion. This equation, that contains 
all the information about the skyrmion-magnon interaction, is written in terms of a 
temperature dependent damping matrix function. In our treatment the possible 
contribution of inelastic process (Cherenkov like radiation) to the soliton dynamics 
is considered. Finally secs. IV and V are devoted to the explicit calculation of the 
damping constant matrix elements that define the skyrmion mobility and to present our 
conclusions. 

\section{Effective Skyrmion-Magnon Hamiltonian}
As we said, the quantum Hall skyrmions can be described in terms of a unit vector 
field associated to the local pseudo-spin orientation as was proposed by Shondi 
\cite{shondi}. The effective lagrangian density of the system can be written as
\begin{equation}
{\mathcal L}({\bf n})=\alpha A\left [ {\bf n} ( {\bf r} )
\right ] \cdot \partial_{t} {\bf n} ( {\bf r} )-V[{\bf n}], 
\label{son}
\end{equation}
where
\begin{equation}
A \left[\bf{n(r)}\right]=\left( \partial _{t}{\bf n}
\times \partial _{\tau } {\bf n}\right)
\end{equation}
is the vector potential of the unit monopole, and the effective potential functional
is that of a ferromagnet with long range interaction arising from the Coulomb repulsion
$V(r)=e^2/r$ between the underlying electrons. 
Explicitly,
\begin{equation}  
V[{\bf n}] =\alpha^{\prime}\left[
\nabla {\bf n} ( {\bf r} ) \right]^{2}+g \overline{\rho} \mu _{B}
{\bf n} ( {\bf r} ) \cdot {\bf B} 
+\frac{1}{2}\int d^{2} r^{\prime}V( {\bf r} - {\bf r}^{\prime}) q({\bf r})
q({\bf r}^{\prime })  \label{azul}   
\end{equation}
where the deviation from the uniform background density ${\overline \rho}=1/2 \pi l_0^2 $ 
is the skyrmion density $q(r)$, which in terms of the $\bf n$ field has the following form
\begin{equation}
q(r)=\frac{1}{8\pi }\varepsilon _{\mu \nu }\mathbf{n}\cdot \left( \partial
_{\mu }\mathbf{n}\times \partial _{\nu }\mathbf{n}\right)
\label{san}.
\end{equation}
Here $\alpha^{\prime}$ is the spin stiffness and $l_0$ stands for the magnetic lenght. 

To take advantage of the internal space spherical symmetry, it is convenient to define the 
${\bf n}$ vector field by means of two escalar fields $\Theta (r,\theta )$ and 
$\Phi (r,\theta )$ such that,
\begin{eqnarray}
n^{x} &=&\sin [\Theta (r,\theta )]\cos [\Phi (r,\theta )]  \nonumber \\
n^{y} &=&\sin [\Theta (r,\theta )]\sin [\Phi (r,\theta )]  \nonumber \\
n^{z} &=&\cos [\Theta (r,\theta )]. \label{tun}
\end{eqnarray}
In terms of these new fields the effective potential functional (\ref{azul}) normalized
to the Zeeman energy reads, 
\begin{eqnarray}
V &=&\alpha ^{\prime }\left[ (\nabla \Theta
)^{2}-\sin ^{2}\Theta \left( \nabla \Phi \right) ^{2}\right]+
g\mu _{B}\overline{\rho }(1-\cos \Theta ) \nonumber \\
&&+\frac{e^{2}}{2} q(r)\left| \frac{\partial (\Theta,\Phi) }{\partial (r,\theta) }\right| 
\int \frac{q(r^{\prime})}{\left| \mathbf{r}-\mathbf{r}^{\prime }\right| }
\left| \frac{\partial (\Theta,\Phi) }{\partial (r^{\prime },\theta ^{\prime })}\right| 
dr^{\prime 2},
\label{seee}
\end{eqnarray}
where $\left| \frac{\partial (\Theta,\Phi) }{\partial (r,\theta)}
\right|$ stands for the Jacobian of the transformation and the skyrmion density is simply 
$q(r)= \sin[ \Theta ]/ 4 \pi r$.\\

From (\ref{seee}), the localized static solutions in spin space (skyrmions) can 
be derived from the variational equations $\delta V \slash \delta \Theta=\delta V 
\slash \delta \Phi=0$. Explicitly,
\begin{equation}
0=\alpha ^{\prime }\left[ \nabla ^{2}\Theta -\frac{\sin 2\Theta }{2}\left(
\nabla \Phi \right) ^{2}\right] -g\mu _{B}B\overline{\rho }\sin \Theta 
+e^{2}q(r) \int q(r^{\prime })\left| \frac{\partial
(\Phi ,\left| \mathbf{r}-\mathbf{r}^{\prime }\right| ^{-1})}{\partial (r,\theta) 
}\right| dr^{\prime 2},  \label{ecu1}
\end{equation}
\begin{equation}
0=\alpha ^{\prime }\left[ \sin ^{2}\Theta \nabla ^{2}\Phi +\sin 2\Theta 
\left(\nabla \Phi \right) .(\nabla \Theta )\right] 
+e^{2}q(r) \int q(r^{\prime })\left| \frac{\partial (\left| \mathbf{r}-\mathbf{r}
^{\prime }\right| ^{-1},\Theta) }{\partial (r,\theta) }\right| dr^{\prime 2}.
\label{ecu2}
\end{equation}
This set of equations with the boundary condition $\Theta \rightarrow 0$ as 
$r \rightarrow \infty $ and cylindrical symmetry was already solved by numerical 
methods \cite{critical}. However, we will use a variational approach, as in 
\cite{lilli}, in which, starting from the non-linear sigma model (NL$\sigma$) solutions 
\cite{poly}
\begin{equation}
\Theta=\arccos\left\{\frac{r^{2Q}-4\lambda^{2Q}}{r^{2Q}+4\lambda^{2Q}}\right\}
\label{pork}
\end{equation}
and
\begin{equation}
\Phi=\arctan\{(\frac{y}{x})^{Q}\},
\label{peek}
\end{equation}
an optimum value of $\lambda$ is determined by balancing the Coulomb and the 
Zeeman contributions in (\ref{seee}). For the topological charge $Q=\int q(r)d^{2}r=1$,
$\lambda$ has the value  
\begin{equation}
\lambda=0.558l_{0}\left ({\widetilde g}\left|\ln\widetilde{g}\right| \right )
^{-\frac{1}{3}}, \qquad {\widetilde g}=\frac{g\mu _{B}B}{ \frac{e^{2}}{l_0} }
\label{taka}
\end{equation}
where $\widetilde{g}$ is the reduced Land\'e factor which varies from 0.02 in
the normal $GaAs$ material to $0.002$ when an hydrostatic pressure is applaied 
\cite{larg}.\\

Therefore, at $T = 0$ and near $\nu=1$, we will assume that the quantum Hall ferromagnet
is a non-interacting skyrmion gas given by (\ref{pork})-(\ref{taka}). However, 
when $T \neq 0$ the ferromagnetic spin wave modes of the system are excited. This kind of 
quantum fluctuations are given essentially by the functional expansion of the 
effective potential $V[\Theta,\Phi]$ 
around the NL$\sigma$ model solution (\ref{pork}) and (\ref{peek}) up to second order. As the 
optimized size for the skyrmion gives a negligible first functional derivative contribution, 
the expansion generates the following equation for the fluctuations of the $\Theta$ field, 
\begin{equation}
\{ \alpha ^{\prime } \nabla ^{2} +U[\Theta,\Phi] \}
u_{m,{\bf k}}({\bf r})=E_{m}({\bf k}) u_{m,{\bf k}}({\bf r}),
\end{equation}
where
\begin{equation}
U[\Theta,\Phi]=\alpha ^{\prime }\cos ( 2\Theta ) (\nabla \Phi)^{2}+g\mu _{B}B
\overline{\rho }\cos \Theta  
+e^{2} q(r) \int q(r^{\prime })K[\Theta,\Phi]dr^{\prime 2}
\label{xixi}
\end{equation}
with
\begin{equation}
K=\left| 
\frac{\partial \left( \Phi(r), \left| \frac{\partial \left( \left| r-r^{\prime}\right|^{-1}
,\Theta(r^{\prime}) \right) }{\partial (r^{\prime}, \theta^{\prime})} \right| \right) }
{\partial (r, \theta)} 
\right|.
\end{equation}
Substituting (\ref{pork}) and (\ref{peek}) in (\ref{xixi}) the following 
Schr\"{o}dinger-like equation, for $Q=1$, is obtained
\begin{equation}
\left\{ -\nabla^{2}+U(r)\right\}
u_{m,{\bf k}}({\bf r})=e_{m}u_{m,{\bf k}}({\bf r}), 
\label{sho}
\end{equation}
where
\begin{equation}
e_{m}=E_{m}-\frac{g\mu_{B}B}{2\pi},
\label{wee}
\end{equation}
and
\begin{equation}
U(r)=\frac{\alpha^{\prime}l^2_0}{r^{2}}-\frac{2^{5} \alpha^{\prime} \lambda ^{2}l^2_0}{(r^{2}+
4\lambda^{2})^{2}}-\frac{2g\mu _{B}B\lambda ^{2}}{\pi(r^{2}+4\lambda ^{2})}.
\label{popo}
\end{equation}

Notice that due to the rotational invariance of (\ref{pork}) and the radial 
symmetry of (\ref{peek}), the Coulomb term does not contribute at all to the 
fluctuations within this approximation, so, far from the scattering centre, the 
fluctuations will be mainly dominated by the potential generated by the skyrmion 
itself and by the presence of the magnetic field.\\

Despite of the complex form of (\ref{popo}), the solutions of (\ref{sho}) for the 
$m$-th cylindrical wave component of the ferromagnetic spin waves confined in a cylindrical 
region of radius $L$ can be written as the sum of two contributions
\cite{morse}
\begin{equation}
u_{mn} \propto \frac{1}{2}\{ H_{|m|}^{(1)}(k_{mn}r) e^{im\phi}+
\sum\limits_{l}
e^{-2i\delta_{m,l}} H_{|l|}^{(2)}(k_{ln}r) e^{il\phi}
\},
\label{funo}
\end{equation}
where $H_{m}^{(1,2)}(kr)$ are the Hankel functions of first and second kind 
respectively and $\delta_{ml}$ is the phase shift matrix which connects the $m$ and 
$l$ angular momentum channels. The first term of the right hand side of (\ref{funo}) 
correspond to the incident cylindrically symmetric wave whereas the second term is 
associated to the shifted $l$-th component after the interaction with the potential 
$U(r)$ has taken place. As can be seen from (\ref{popo}), the potential involved in the 
scattering processes is cylindrically symmetric, therefore $\delta_{m,l}$ is a diagonal 
matrix and (\ref{funo}) becomes
\begin{equation}
u_{mn} \propto \frac{1}{2}\{ 
H_{|m|}^{(1)}(k_{mn}r) +
e^{-2i\delta_{m}} H_{|m|}^{(2)}(k_{mn}r) \}e^{im\phi}.
\label{simple}
\end{equation}

To calculate the phase shifts for each angular momentum component we used the Fredholm method
\cite{fred}. In so doing it is convenient to start from the well-known relation,
\begin{equation}
\pi { \cal A}(E)\cot \delta (E)=1+{\cal P}\int\limits_{0}^{\infty }
\frac{{ \cal A}(E^{\prime})}{E-E^{\prime}}dE^{\prime },
\label{fff}
\end{equation}
where ${\cal P}$ means the Cauchy principal value, and ${ \cal A}(E)$
can be obtained in first approximation as
\begin{equation}
{ \cal A}(E)=
-\langle E |U| E \rangle.
\label{ae}
\end{equation}
In computing  ${ \cal A}(E)$ we will use a cylindrical basis solution of the non 
perturbed Schr\"{o}dinger problem with circular boundary of radius $L$. When $L$ is 
large enough, the 2-D cylindrical basis can be written as
\begin{equation}
\langle r| E \rangle =\left(\frac{2M}{\hbar^{2}} \right)^{\frac{1}{2}}
J_{m}(kr) e^{i m \theta}.
\label{ai}
\end{equation}

However looking at (\ref{popo}) it can be seen that, as $r \rightarrow 0$, $U(r) 
\rightarrow \infty$ and hence every component of (\ref{ai}) must vanish at $r=0$.
This condition is automatically satisfied by the $m \neq 0$ $J_m(kr)$ functions. 
Therefore substituting (\ref{ai}) into (\ref{ae}) we get
\begin{eqnarray}
{ \cal A}_{m}(x)&=&\frac{\pi x e_{c}}{8 \sqrt {2\pi}} 
\left\{ I_{m}(x) [K_{m+1}(x)+K_{m-1}(x)]  
-K_{m}(x)[I_{m+1}(x)+I_{m-1}(x)] - (4mx)^{-1} \right\} \nonumber \\
&&+ 2 e_{z} I_{m}(x) K_{m}(x),
\label{aene}
\end{eqnarray} 
with
\begin{equation}
e_{c}=\frac{e^2/l_0}{\hbar^2/M l^2_0}, \qquad e_{z}=\frac{g \mu_B B}{\hbar^2/M\lambda^2},
\end{equation}
where we have used the value of the spin stiffness $e^2/32\sqrt{2\pi} l_0$ of the quantum 
Hall ferromagnet. On the other hand $I_{m}$ and $K_{m}$ are the modified Bessel fuctions 
and $x=2k\lambda$. Then, the phase shift for each $m$ angular momentum component is given by
\begin{equation}
\delta_{m}=\arctan\left(\frac{\pi{\cal A}_{m}(2 \lambda k)}{1+\Delta_{m}(2 \lambda k)}
\right),
\label{ppp}
\end{equation}
where we have used $\Delta(2 \lambda k)$ to denote the expression
\begin{equation}
\Delta_{m}(k)=
{\cal P} \int\limits_{0}^{\infty } \frac{{ \cal A}_{m}(x)}{(2\lambda k)^{2}-x^{2}} xdx.
\label{phase}
\end{equation}

So, until now, our model for the quantum Hall ferromagnet is composed by a set of 
static cylindrically symmetric skyrmion excitations given by (\ref{pork})-(\ref{taka})
and non interacting spin waves modes described by (\ref{simple}) and 
(\ref{aene})-(\ref{phase}) with frequencies  
\begin{equation}
\omega _{mn}=\frac{ \hbar k_{mn}^{2} }{2M}+ \frac {g\mu_{B}B}{\hbar},
\label{fri}
\end{equation}
where the momenta $k_{mn}$ can be determined from the boundary condition $u_{mn}(kL)=0$. 
Therefore using the asymptotic expansion of (\ref{simple}) we get 
\begin{equation}
k_{mn}=\frac{2n+1}{2L}+\frac{m \pi}{2L}-\frac{\delta_{m}}{L}.
\end{equation}

On the classical level, due to the translational invariance of the system, the skyrmions 
can move freely, and the position of its centre of mass will be enough to describe its 
motion. However, if we look at the problem from the quantum field theoretical point of 
view, the skyrmion will be a particle and its centre of mass a dynamical
variable. In this picture, due to the excitation of the spin wave modes at finite $T$, 
not all of the degrees of freedom of the system will contribute to the skyrmion formation, 
this can be interpreted as the origin of the  residual skyrmion-magnon interaction that 
leads to a non trivial dynamics of these objects. To show that, we explicitly evaluate the 
equation of motion for the $\Theta$ field (our skyrmion field now) from the equations 
(\ref{son}),(\ref{tun}) and (\ref{seee}). Considering cylindrical symmetry for non static 
skyrmions and neglecting higher order spatial derivatives, the effective 
equation of motion for $\Theta ({\bf r},t)$ can be written as,
\begin{equation}
\frac{1}{c^{2}}\stackrel{\cdot \cdot }{\Theta }+2\alpha ^{\prime }\left[
\nabla ^{2}\Theta -\frac{\sin 2\Theta }{2}\right] -g\overline{\rho }\mu
_{B}B\sin \Theta =0, \label{fio}
\end{equation}
where $c=\alpha ^{\prime }\slash{\alpha }$. Comparing (\ref{ecu1}) and 
(\ref{fio}) it can be seen that to describe the skyrmion dynamics interacting with a magnon 
thermal bath in a simple way, we can start from the one field Lagrangian density associated 
with the ${\bf z}$ pseudo-spin component which reads
\begin{equation}
{\cal L}=\frac{1}{2c^{2}}\left( \stackrel{\cdot }{\Theta }\right)
^{2}-V[\Theta ], \label{suki}
\end{equation}
where
\begin{equation}
V[\Theta ]=\alpha ^{\prime }\left[ (\nabla \Theta
)^{2}-\frac{\sin ^{2}\Theta }{r^{2}}\right] 
+g\mu _{B}B\overline{\rho }(1-\cos \Theta ).
\label{teta}
\end{equation}

Therefore, our assumptions have simplified the initial problem given by (\ref{son})-(\ref{san}) 
into an effective two dimensional one escalar field theory that describes the 
dynamical skyrmion and the possible scattering states given by ferromagnetic spin waves.\\

At this point, our model to treat the transport properties of the skyrmions is given by
(\ref{suki}) and (\ref{teta}). The correct quantization of (\ref{suki}) can be done using 
the canonical formalism via the method of collective coordinates \cite{christlee}. Doing this
we approprietely treat the zero frequency mode associated to the translational invariance, and 
the second quantized version of $\widehat{H}$ (obtained from (\ref{suki}) and (\ref{teta})) 
which describes the momentum-momentum coupling between skyrmions and magnons (see \cite{pola}
for details) can be written as
\begin{equation}
\widehat{H}=\frac{1}{2M}\left( \widehat{{\bf P}}_{s}-\widehat{{\bf P}}_{mg}\right) ^{2}+
\sum\limits_{mn}
\hbar \omega _{mn}b_{mn}^{\dagger }b_{mn}.  
\label{aecn}
\end{equation}

In (\ref{aecn}), $\widehat{{\bf P}}_{s}$ and $\widehat{{\bf P}}_{mg}$ denotes the skyrmion and magnon 
momentum operators and $\omega_{mn}$ is given by (\ref{fri}). Explicitly,
\begin{equation}
\widehat{{\bf P}}_{mg}=\sum\limits_{mn,kl}{\bf D}_{mn,kl} 
b_{mn}^{\dagger }b_{kl} 
+\sum\limits_{mn,kl}{\bf C}_{mn,kl} \left( b_{mn}b_{kl}-b_{mn}^{\dagger
}b_{kl}^{\dagger }\right),  \label{aecn1}
\end{equation}
with
\begin{equation}
{\bf D}_{mn,kl}=\frac{1}{2}
\left[ \sqrt{\left( \frac{\omega _{kl}}{\omega _{mn}}\right) }+\sqrt{
\left( \frac{\omega _{mn}}{\omega _{kl}}\right) }\right]{\bf G}_{mn,kl},
\label{fdos}
\end{equation}
\begin{equation}
{\bf C}_{mn,kl}=\frac{1}{4}\left[ \sqrt{\left( 
\frac{\omega _{kl}}{\omega _{mn}}\right) }-\sqrt{\left( \frac{\omega _{mn}}{%
\omega _{kl}}\right) }\right]{\bf G}_{mn,kl},
\label{lsd}
\end{equation}
where ${\bf G}_{mn,kl}$ is a $2$-D vector given by
\begin{equation}
{\bf G}_{mn,kl}=\int u_{kl}({\bf r}) \nabla u_{mn}({\bf r}) d^{2}r
\label{tenso}
\end{equation}
and
\begin{equation}
M=\int (\nabla u_{0}({\bf r}))^2 d^{2}r,
\label{impotta}
\end{equation}
where $u_0$ denote the localized solution (\ref{pork}) with $Q=1$ whose gradient is
associated to the zero frequency mode.

The relations (\ref{aecn})-(\ref{impotta}) are a generalization for the 2-D case of the 
results obtained in \cite{pola}. Notice from (\ref{aecn1}), that the magnon momentum 
operator consists of two contributions. A block diagonal term, which only couples excitations
with different $k$, as can be seen from the antisymmetric behaviour of ${\bf D}_{mn,kl}$ under
inversion of index, and that is responsible for the low energy scattering of magnons by the 
skyrmions. This term commutes with the total number of magnons, so any kind of 
transition involving processes which do not conserve the number of magnons is not 
allowed. Actually, this contribution remind us of the problem of scattering by a hard 
(but not fixed) object because there are no internal degrees of freedom in our model.
On the other hand, the block off-diagonal terms in (\ref{aecn1}) describes the emission or 
absorption of magnons by the skyrmion. In some specific situations, during the Hall 
measurements or in the high temperature limit, for instance, the effective motion of the 
skyrmion can be affected by the off-diagonal contribution. Although we do not attempt to 
study the contribution of the inelastic process  to the skyrmion dynamics in this paper, we 
will show that in the long time approximation this term do not contribute at all to the 
transport properties.

\section{The Feynman-Vernon Formalism}
Based on the functional integral method of the Feynman-Vernon formalism 
\cite{feynman}, Castro Neto and Caldeira developed a new model for dissipation in quantum 
mechanics \cite{lett} to study the polaron dynamics \cite{pola} without appealling 
to the kinetic theory. In this work we will follow the same kind of approach, but now in the 
2-D physical space, in our study of the effective skyrmion dynamics in the interacting 
system (skyrmion plus magnons). \\

The key point of the method is always to compute the reduced density operator 
($\widehat{\rho}_{s}$) for the particle of interest (the skyrmion) by tracing the mesons
(magnons) coordinates out of the problem. The density operator of the interacting system 
evolves in time according to
\begin{equation}
\widehat{\rho }(t)=e^{-\frac{i}{\hbar}\widehat{H}t} \widehat{\rho }
(0) e^{\frac{i}{\hbar}\widehat{H}t},
\end{equation}
where $\widehat{H}$ is given by (\ref{aecn}) and $\widehat{\rho}(0)$ is assumed to be 
decoupled at $t=0$ as
\begin{equation}
\widehat{\rho}(0)=\widehat{\rho}_{s}(0) \widehat{\rho}_{mg}(0),
\end{equation}
where ${\underline s}$ and ${\underline mg}$ denotes skyrmion and magnon respectively and
\begin{equation}
\widehat{\rho}_{mg}(0)=\frac{e^{-\beta \widehat{H}_{mg}}}
{Tr(e^{-\beta \widehat{H}_{mg}})}.
\end{equation}

As it was shown in \cite{pola}, the reduced density operator in the coordinate representation 
for the particle of interest (skyrmion) can be written as
\begin{equation}
\widehat{\rho }({\bf x},{\bf y},t)=\int \int d{\bf x}^{\prime }d{\bf y}^{\prime} 
{\mathcal J}({\bf x},{\bf y},t;{\bf x}^{\prime},{\bf y}^{\prime },0)\widehat{\rho }_{s}
({\bf x}^{\prime },{\bf y}^{\prime},0),
\label{bolllo}
\end{equation}
where ${\cal J}$ is the skyrmion superpropagator which can be explicitly written in terms 
of the influence functional ${\cal F}[{\bf x},{\bf y}]$ as
\begin{equation}
{\mathcal J}=\int\limits_{{\bf x}^{\prime}}^{{\bf x}}
{\mathcal D} {\bf x}\int\limits_{{\bf y}^{\prime }}^{{\bf y}} {\cal D}{\bf y} 
e^ {\frac{i}{\hbar}(S_{0}[{\bf x}]-S_{0}[{\bf y}])}{\mathcal F}[{\bf x},{\bf y}],
\label{sucre}
\end{equation}
where the action associated to the free skyrmion motion is
\begin{equation}
S_{0}\left[ {\bf x} \right] =\int\limits_{0}^{t}M\frac{(\stackrel{.} {\bf x}
)^{2}}{2}dt,  \label{sop}
\end{equation}
with $M$ given by (\ref{impotta}), and the influence funtional in the coherent 
state representation is written as
\begin{eqnarray}
{\mathcal F}&=&\int \frac{d^{2} \alpha }{\pi }\int \frac{d^{2}\beta }{
\pi }\int \frac{d^{2}\gamma }{\pi }
\left\{\right.
e^{-\left| \alpha \right| ^{2}-\frac{\left| \beta \right|^{2}}{2}
-\frac{\left|  \gamma \right| ^{2}}{2}} 
\rho _{mg}\int\limits_{\beta }^{{\alpha }^{*}}{\mathcal D}\alpha
\int\limits_{{\gamma }^{*}}^{\alpha}{\mathcal D} \gamma
e^{\frac{i}{\hbar }\left\{ S_{I}[{\bf x},{\bf \alpha }]-S_{I}[{\bf y},{\bf \gamma}]
\right\}} \times \nonumber \\  
&&e^{({\bf \alpha}^{*}(0){\bf \alpha}+{\bf \alpha}^{*}
{\bf \alpha}(t)+{\bf \gamma }(0){\bf \alpha }^{*}+
{\bf \alpha \gamma }^{*}(t))}
\left.\right\}.  
\label{inl}
\end{eqnarray}
Here we have introduced $S_{I}$ as the interacting skyrmion-magnon action, and
$\rho _{mg}$ stands for the magnon density operator in the coherent state 
representation,
\begin{eqnarray}
S_{I}&=&\int\limits_{0}^{t}\left\{\right.
\frac{i\hbar }{2}
\sum\limits_{mn} \alpha _{mn}^{*}\stackrel{.}{\alpha }_{mn}-\alpha _{n}
\stackrel{.}{\alpha }_{mn}^{*}-\sum\limits_{mn}\hbar
\omega _{mn}\alpha _{mn}^{*}\alpha _{mn}\nonumber \\
&&+\hbar M \stackrel{.}{{\bf x}} \cdot \sum\limits_{mn,kl}{\bf D}_{mn,kl}\alpha _{mn}^{*}
\alpha _{kl} \nonumber \\
&&+\hbar M \stackrel{.}{{\bf x}} \cdot \sum\limits_{mn,kl}{\bf C}_{mn,kl}(\alpha _{mn}
\alpha _{kl}+
\alpha _{mn}^{*}\alpha _{kl}^{*})\left.\right\}dt^{\prime},
\label{sip}
\end{eqnarray}
and
\begin{equation}
\rho _{mg}(\beta ^{*},\gamma )=\prod\limits_{mn}\frac{e^{ \beta
_{mn}^{*}\gamma _{mn}e^{-\beta \hbar \omega _{mn}}-\frac{\left| \beta
_{mn}\right| ^{2}}{2}-\frac{\left| \gamma _{mn}\right| ^{2}}{2}}}
{(1-e^{-\beta \hbar \omega _{mn}})^{-1}}.
\end{equation}

Notice from (\ref{sip}) that changing to the Lagrangian formalism through the coherent 
state representation we have simplified the problem (\ref{aecn}), obtaining an 
interacting skyrmion-magnon action quadratic in ${\bf \alpha}$. Therefore 
the equation of motion obtained from the variational problems 
$\delta S_{I} \slash \delta \alpha=\delta S_{I} \slash \delta \alpha^{*}=0$ 
can be formally solved. \\

The linear Euler-Lagrange equations obtained from (\ref{sip}) can be written as
\begin{equation}
\stackrel{.}{\alpha }_{mn}+i \omega _{mn}\alpha _{mn}-i M \stackrel{.}{{\bf x}} \cdot
\sum\limits_{kl}{\bf D}_{mn,kl}\alpha _{kl} 
+2{\bf C}_{mn,kl}\alpha_{kl}^{*}=0,  
\label{uno1}
\end{equation}
\begin{equation}
\stackrel{.}{\alpha }_{mn}^{*}-i \omega _{mn} \alpha _{mn}^{*}+i M
\stackrel{.}{{\bf x}} \cdot \sum\limits_{kl}{\bf D}_{mn,kl}\alpha _{kl}^{*} 
+2{\bf C}_{mn,kl} \alpha_{kl}=0,  
\label{dos2}
\end{equation}
whose solutions, in terms of the functional operator $W_{nm}$ are of the form, 
\begin{equation}
\alpha _{mn}(\tau )=e^{-i\omega_{nm}\tau}\left[\alpha _{nm}+\sum\limits_{m\neq n}W
_{nm}(\tau )\alpha _{m}\right], \label{alfa1}
\end{equation}
\begin{equation}
\alpha _{n}^{*}(\tau )=e^{i\omega_{n}\tau}\left[\alpha _{n}^{*}+\sum\limits_{mn\neq 
kl}\left(W_{mn,kl}(\tau )\alpha _{kl}\right) ^{*}e^{-i\omega_{kl}t}\right]. 
\label{alfa2}
\end{equation}
The functional operator introduced above obeys the integral equation 
\begin{eqnarray}
W_{mn}[\bf{x}(\tau)] &=&\int\limits_{0}^{\tau }W_{mn,kl}^{0}
(\tau ^{\prime })d\tau ^{\prime } \nonumber \\
&&+\int\limits_{0}^{\tau}\sum\limits_{pq\neq mn} W_{mn,pq}^{0}(\tau ^{\prime })W_{pq,kl}
(\tau ^{\prime })d\tau ^{\prime }  \label{flip2}
\end{eqnarray}
with
\begin{equation}
W_{mn,kl}^{0}=iM\stackrel{.}{{\bf x}} \cdot \left\{{\bf D}_{mn,kl}e^{-i(\omega_{mn}-
\omega_{kl})t} 
+ 2{\bf C}_{mn,kl}^{*}e^{i(\omega_{mn}+\omega_{kl})t} \right\}.
\label{flip3}
\end{equation}

Now, using (\ref{alfa1}) and (\ref{alfa2}) we are able to solve the functional integrals 
in (\ref{inl}) in the stationary phase approximation. After evaluating some integrals, the 
influence functional can be further expressed as
\begin{equation}
F[{\bf x},{\bf y}]=\frac {1}{\det \left(1-\overline{N}_{mn}{\large
\Gamma }_{mn,kl}\right) },  
\label{xxs}
\end{equation}
where $\Gamma_{mn,kl}$ is given by
\begin{equation}
{\large \Gamma }_{mn,kl}=W_{mn,kl}[{\bf y}]+ W_{mn,kl}^{\dagger }[{\bf x}]+
W_{mn,pq}[{\bf y}] W_{pq,kl}^{\dagger }[{\bf x}].  \label{ssm}
\end{equation}
and the ocupation number for magnons is,
\begin{equation}
\overline{N}_{mn}=\frac{1}{\exp (\beta \hbar
\omega _{mn})-1}.
\label{numero}
\end{equation}

To go on in computing the influence functional we need to explicitly solve (\ref{flip2}). 
This can be done up to any order by iteration, however as this relation is nothing but the  
amplitude of scattering from one state $mn$ to another $kl$ through virtual transitions 
involving the state $pq$, we hope that in describing the low velocity skyrmion dynamics
only small energy processes are involved and therefore a second order iteration will 
be sufficient to describe the main features of the transport properties of the system. 
Doing this, and after some algebra, the influence functional can be expressed as
\begin{equation}
F[{\bf x},{\bf y}]=e^{\frac{i}{\hbar }(\Phi ^{I}+\Psi ^{I})}
e^{-\frac{1}{\hbar }(\Phi ^{R}+\Psi ^{R})}.
\label{qoqo}
\end{equation}
Where the following definitions were used,
\begin{equation}
\Phi ^{R}=\int\limits_{0}^{t}dt^{\prime }\int\limits_{0}^{t^{^{\prime
}}}dt^{\prime \prime } \sum _{i,j=1}^{2} \left \{
\left[ \stackrel{.}{x}_i(t^{\prime })+\stackrel{.}{y}_i
(t^{\prime })\right] 
\Gamma _{i,j}^{R}(t^{\prime }-t^{\prime \prime }) \left[ 
\stackrel{.}{ x}_j(t^{\prime \prime })-\stackrel{.}{ y}_j(t^{\prime \prime })
\right] \right \},
\end{equation}
\begin{equation}
\Phi ^{I}=\int\limits_{0}^{t}dt^{\prime }\int\limits_{0}^{t^{^{\prime
}}}dt^{\prime \prime } \sum _{i,j=1}^{2} \left \{ 
\left[ \stackrel{.}{x}_i(t^{\prime })+\stackrel{.}{%
y}_i(t^{\prime })\right] 
\Gamma _{i,j}^{I}(t^{\prime }-t^{\prime \prime })
\left[ \stackrel{.}{ x}_j(t^{\prime \prime })-\stackrel{.}{ y}_j(t^{\prime \prime })
\right] \right \} ,
\end{equation}
\begin{equation}
\Psi ^{R}=\int\limits_{0}^{t}dt^{\prime }\int\limits_{0}^{t^{^{\prime
}}}dt^{\prime \prime } \sum _{i,j=1}^{2}
\left[ \stackrel{.}{ x}_i(t^{\prime \prime})\Delta _{i,j}^{R}(t^{\prime }-t^{\prime \prime }) 
\stackrel{.}{ y}_j(t^{\prime})
-\stackrel{.}{ x}_i(t^{\prime})\Delta _{i,j}^{R}(t^{\prime }-t^{\prime \prime }) 
\stackrel{.}{ y}_j(t^{\prime \prime})\right] 
\end{equation}
\begin{equation}
\Psi ^{I}=\int\limits_{0}^{t}dt^{\prime }\int\limits_{0}^{t^{^{\prime
}}}dt^{\prime \prime } \sum _{i,j=1}^{2}
\left[ \stackrel{.}{x}_i(t^{\prime \prime}) \Delta _{i,j}^{I}(t^{\prime }-t^{\prime \prime }) 
\stackrel{.}{y}_j(t^{\prime}) 
+ \stackrel{.}{x}_i(t^{\prime}) \Delta _{i,j}^{I}(t^{\prime }-t^{\prime \prime }) 
\stackrel{.}{y}_j(t^{\prime \prime})
\right]
\end{equation}
where $\Gamma_{i,j}$ and $\Delta _{i,j}$ are $2 \times 2$ matrices with elements involving 
the components of the $2$-D vectors ${\bf D}_{mn,kl}$ and ${\bf C}_{mn,kl}$ given by 
(\ref{fdos})-(\ref{tenso}). Explicitly
\begin{equation}
\Gamma _{i,j}^{R}(t)=\sum\limits_{mn,kl} 
\frac{(\overline{N}_{mn}+\overline{N}_{kl}+2\overline{N}_{mn}\overline{N}_{kl})
(\omega_{mn} +\omega_{kl})^2}{2 \omega_{mn} \omega_{kl}} 
{\cal M}_{i,j}(mn,kl) \cos (\omega_{kl} - \omega_{mn})t ,
\end{equation}
\begin{equation}
\Gamma _{i,j}^{I}(t)=\sum\limits_{mn,kl}\frac{(\overline{N}_{mn}-\overline{N}_{kl})
(\omega_{mn} +\omega_{kl})^2}{2 \omega_{mn} \omega_{kl}}
{\cal M}_{i,j}(mn,kl) \sin (\omega_{kl} - \omega_{mn})t ,
\label{cinco}
\end{equation}
\begin{equation}
\Delta _{i,j}^{R}(t)=\sum\limits_{mn,kl} 
\frac{(\overline{N}_{mn}+\overline{N}_{kl}+2\overline{N}_{mn}\overline{N}_{kl})
(\omega_{mn} - \omega_{kl})^2}{2 \omega_{mn} \omega_{kl}} 
{\cal M}_{i,j}(mn,kl) \cos (\omega_{kl} + \omega_{mn})t ,
\end{equation}
\begin{equation}
\Delta _{i,j}^{I}(t)=\sum\limits_{mn,kl} 
\frac{(\overline{N}_{mn}+\overline{N}_{kl}+2\overline{N}_{mn}\overline{N}_{kl})
(\omega_{mn} - \omega_{kl})^2}{2 \omega_{mn} \omega_{kl}}
{\cal M}_{i,j}(mn,kl) \cos (\omega_{kl} + \omega_{mn})t,
\label{seis}
\end{equation}
\begin{eqnarray}
{\cal M}_{i,j}&=&\left(
\begin{array}{cc}
|G_{mn,kl}^{(1)}|^2 & 0 \\
0 & |G_{mn,kl}^{(2)}|^2 
\end{array}
\right)
\label{matrix}
\end{eqnarray}
where $G_{mn,kl}^{(1),(2)}$ are the components of the ${\bf G}$ vector given by (\ref{tenso}).
Finally, substituting (\ref{qoqo}) in (\ref{sucre}) the skyrmion superpropagator becomes
\begin{equation}
{\cal J}=\int\limits_{{\bf x}^{\prime }}^{{\bf x}}
{\mathcal D}{\bf x}\int\limits_{{\bf y}^{\prime }}^{{\bf y}}{\mathcal D}{\bf y}e^ 
{i \overline{S}[{\bf x},{\bf y}]+ i \hbar(\Phi ^{R}-\Psi ^{R})}.
\label{fizz}
\end{equation}

Now, after having traced the magnons coordinates out, the reduced skyrmion action contains 
all the information of the interaction with the thermal bath and can be written in a simple 
way as
\begin{equation}
\overline{S}[{\bf x},{\bf y}]=S_{0}[{\bf x}]-S_{0}[{\bf y}]+i\hbar(\Phi ^{I}+\Psi^{I}).
\label{manca}
\end{equation}

Before analyzing effective skyrmion motion it is convenient to rewrite
(\ref{manca}) in terms of a new set of variables, ${\bf R}=({\bf x}+{\bf y})/2$ and 
${\bf r} =({\bf x}-{\bf y})$. The equations 
of motion associated to the action  $\overline{S}[{\bf R},{\bf r}]$ can be written as
\begin{equation}
\stackrel{\cdot \cdot }{R} _{i}(t) + \int\limits_{0}^{t}\gamma _{i,j} (t-t^{\prime })%
\stackrel{\cdot }{R}_j (t^{\prime })dt^{\prime }=0, 
\label{em1}
\end{equation}
\begin{equation}
\stackrel{\cdot \cdot }{r}_{i} (t) -\int\limits_{0}^{t}\gamma _{i,j} (t-t^{\prime })%
\stackrel{\cdot }{r}_j (t^{\prime })dt^{\prime }=0, 
\label{em2}
\end{equation}
where we have introduced the matrix damping function, $\gamma_{i,j} (t-t^{\prime})$, that
characterize in a general form the mobility of the $2$-D solitons. In terms of (\ref{cinco}),
(\ref{seis})  and (\ref{matrix}) the matrix $\gamma_{i,j}$ is given by
\begin{equation}
\gamma_{i,j} (t-t^{\prime})=-\frac{2\hbar }{M}\frac{d}{dt} \left\{
\Gamma _{i,j}^{I}(t-t^{\prime})
+\Delta _{i,j}^{I}(t-t^{\prime}) \right\}. 
\label{xita}
\end{equation}

The equations of motion (\ref{em1}) and (\ref{em2}) are generalizations of those 
obtained in the case of the soliton dynamics \cite{lett} describing a quantum Brownian 
motion with memory effect. The main differences are that the present treatment allows for the 
study of the $2$-D topological excitations and that the Cherenkov term accounts for the 
possibility of mesons (magnons) emission or absorption by the soliton (skyrmion).
This contribution can be important, even for small velocities, if we are studying the
time evolution of the soliton. 

On the other hand,  as in \cite{lett}, the real part of the exponent in (\ref{manca}) is 
related to the matrix diffusion function 
\begin{equation}
D_{i,j}(t)=\hbar \frac{d^{2}}{dt^{2}}(\Gamma_{i,j}^{R}+\Delta_{i,j}^{R}),
\end{equation}
that can be computed in close analogy to the matrix damping functions as shown 
in the next section.\\

To simplify the analysis of (\ref{xita}) the explicit expression for the damping 
function will be written as a sum of two terms, 
\begin{equation}
\gamma_{i,j} (t-t^{\prime})=\gamma_{i,j}^{E} (t-t^{\prime})+\gamma_{i,j}^{I} (t-t^{\prime}),
\end{equation}
where
\begin{equation}
\gamma_{i,j}^{E}=\frac{\hbar }{M}  \sum\limits_{mn,kl}
\frac{(\overline{N}_{mn}-\overline{N}_{kl})
(\omega_{mn} +\omega_{kl})^2(\omega_{kl}-\omega_{mn})}
{2 \omega_{mn} \omega_{kl}}
{\cal M}_{i,j}(mn,kl) \cos (\omega_{kl}-\omega_{mn})(t-t^{\prime}) ,
\label{cris} 
\end{equation}
and
\begin{eqnarray}
\gamma_{i,j}^{I}&=&\frac{2\hbar }{M}\sum\limits_{mn,kl}
\frac{(\overline{N}_{mn}+\overline{N}_{kl}+2\overline{N}_{mn} \overline{N}_{kl})}
{ \omega_{mn} \omega_{kl}} {\cal M}_{i,j}(mn,kl) \times \nonumber \\
&&(\omega_{mn} -\omega_{kl})^2(\omega_{kl}+\omega_{mn}) 
\cos (\omega_{kl}+\omega_{mn})(t-t^{\prime}).
\label{cristi}
\end{eqnarray}

As it can be seen from (\ref{tenso}) and (\ref{matrix}) the matrix elemements 
${\cal M}_{i,j}$ essentially involve the single particle momentum operator
between two different ferromagnetic spin wave states given by (\ref{simple}). Therefore
in close analogy to the spectral fuctions introduced in \cite{cita} we will define  
the $2 \times 2$ matrix scattering function, $S_{i,j}$, as
\begin{equation}
S_{i,j}(\omega ,\omega ^{\prime })=\sum\limits_{mn,kl} {\cal M}_{i,j}(mn,kl)
\delta(\omega -\omega _{mn})\delta (\omega ^{\prime }-\omega _{kl}),
\label{scatt}
\end{equation}
that allow us to rewrite the expressions for $\gamma_{i,j}^{I}$ and $\gamma_{i,j}^{E}$ in
the following form
\begin{eqnarray}
\gamma_{i,j}^{E} (t)&=&\frac{\hbar }{M} \Theta(t) \int d\omega \int d\omega^{\prime} 
S_{i,j}(\omega, \omega^{\prime})(\overline{N}(\omega)-\overline{N}(\omega^{\prime}))
\nonumber \\
&&\frac{(\omega +\omega^{\prime})^2(\omega^{\prime}-\omega)}
{ \omega \omega^{\prime}} \cos (\omega^{\prime}-\omega)t ,
\label{salute} 
\end{eqnarray}
which corresponds to the elastic scattering process with purely instantaneous memory when 
we assume long time approximation, as in \cite{pola}. On the other hand
\begin{eqnarray}
\gamma_{i,j}^{I} (t)&=&\frac{2\hbar }{M} \Theta(t) \int d\omega \int d\omega^{\prime} 
\frac{S_{i,j}(\omega, \omega^{\prime})(\omega - \omega^{\prime})^2(\omega^{\prime}+\omega)}
{ \omega \omega^{\prime}} \nonumber \\
&&[ \overline{N}(\omega)+\overline{N}(\omega^{\prime})
+2\overline{N}(\omega)\overline{N}(\omega^{\prime})] \cos (\omega +\omega^{\prime})t,
\label{war}
\end{eqnarray}
is related to the Cherenkov radiation and only contributes to the soliton dynamics  
when transitions between states with different energies are taken into account.  
So, we will refer to this contribution as inelastic. In (\ref{salute}) and (\ref{war}) 
$\Theta(t)$ ensures the causality principle.

\section{The Damping Matrix}
In the first place let us analyze the $\gamma_{i,j}^{I} (t)$ contribution to the damped 
motion of the skyrmion. Notice from (\ref{war}) that for long times the 
$\cos(\omega^{\prime}+\omega)t$ term oscillates rapidly giving no net contribution to 
the damping function matrix elements. This is true unless  $\omega^{\prime}+\omega$ is 
close to zero, but at the same time the factor $(\omega -\omega^{\prime})^2$ prevents this 
situation. So, up to first order in the solution of (\ref{flip2}), the inelastic contribution
$\gamma_{i,j}^{I} (t)$ to the damping function can be neglected. However, if we consider 
processes with characteristic times of the same order as $(\omega^{\prime}+\omega)^{-1}$,
when the long time approximation is no longer valid, this term will play an 
important role in the soliton dynamics. \\

Now, the analysis of (\ref{salute}) becomes simpler if we change the frequency variables 
to $\varphi= (\omega ^{\prime }+\omega)/2$ and $\eta= (\omega ^{\prime }-\omega)$. In this way
\begin{equation}
\gamma_{i,j}^{E}(t)=\frac{2\hbar }{M} \Theta(t) \int\limits_{0}^{\infty} d\varphi 
\int\limits_{-\infty}^{\infty} d\eta 
S_{i,j}(\varphi,\eta) {\eta}^{2} f(\varphi, \eta) 
\frac{{\overline N} (\varphi + \eta /2 )-{\overline N}(\varphi - \eta /2 )}{\eta}
\cos (\eta t).
\label{bindi} 
\end{equation}
where
\begin{equation}
f(\varphi, \eta)=2+\frac{\varphi + \eta /2}{\varphi - \eta /2}+
\frac{\varphi - \eta /2}{\varphi + \eta /2}.
\end{equation}

As in (\ref{war}), the expression (\ref{bindi}) vanishes for long times due to the 
rapid oscillations of the $\cos (\eta t)$ term unless $\eta$ is close to zero. 
If this is so $f(\varphi , \eta)$ can be taken as constant and we can assume that
\begin{equation}
\frac{{\overline N} (\varphi + \eta /2 )-{\overline N}(\varphi - \eta /2 )}{\eta}
\approx \frac{\partial \overline{N}(\varphi)}{\partial \varphi},
\end{equation}
therefore (\ref{bindi}) can be rewritten as
\begin{equation}
\gamma_{i,j}^{E} (t)=\frac{8\hbar }{M} \Theta(t) \int\limits_{0}^{\infty} d\varphi 
\int\limits_{-\infty}^{\infty} d\eta A_{i,j}( \varphi )  
\frac{\partial \overline{N}(\varphi)}{\partial \varphi}
\cos (\eta t), 
\label{sangre}
\end{equation}
where we have introduced $A_{i,j}(\varphi)$ given by
\begin{equation}
A_{i,j}(\varphi)=\lim_{\eta \rightarrow 0} S_{i,j}(\varphi,\eta) \eta ^{2}.
\label{chuva}
\end{equation}

After integrating (\ref{sangre}) in $ d \eta $ and using (\ref{numero}) we get 
\begin{equation}
\gamma_{i,j}^{E} (t)=\overline {\gamma}_{i,j}(T) \delta(t),
\label{mark}
\end{equation}
where $\overline {\gamma}_{i,j}(T)$ is the temperature dependent matrix damping parameter
that characterizes the soliton mobility in the $2$-D case and $\delta (t)$ is the Dirac 
delta function. As usual the we will refer to the form (\ref{mark}) as the Markovian 
approximation, because this results have only instantaneus memory not depending on the previous
motion of the particle. Explicitly $\overline {\gamma}_{i,j}(T)$ is given by
\begin{equation}
\overline {\gamma}_{i,j}(T)=\frac{8\hbar}{M}\int\limits_{0 }^{\infty }\frac
{e^{\beta(g\mu _{B}B+\hbar\varphi)}}
{(1-e^{\beta(g\mu _{B}B+\hbar \varphi)})^2}A_{i,j}(\varphi)d\varphi.
\label{land}
\end{equation}

In order to compute the temperature dependence of ${\overline \gamma}_{i,j}(T)$ we need to 
explicitly evaluate the scattering function $S_{i,j}(\varphi,\eta)$. The first step in doing 
so is to get an analitic expression for $S_{i,j}(\omega,\omega^{\prime})$, and therefore the 
matrix elements of (\ref{matrix}) must be evaluated. As it can be seen from (\ref{simple}) the 
{\bf x} and {\bf y} components of the ${\bf G}$ vector defined by (\ref{matrix}) are equal, 
in agreement with the isotropic character of our model. Therefore the function matrix 
${\cal M}_{mn,kl}$ can be written as ${\cal M}=G_{mn,kl}{\bf 1}$ where ${\bf 1}$ is the 
unitary matrix. Then we need to compute only one of its componenets, for instance,
\begin{eqnarray}
G_{mn,kl}&=&\int\limits_{0}^{L}rdr\int\limits_{0}^{2\pi}d\theta u^{*}_{kl}(\cos\theta \frac{\partial}
{\partial r}-\frac{\sin\theta}{r}\frac{\partial}{\partial \theta})u_{mn}
\end{eqnarray} 
which, after some integration and assuming large values of $L$ becomes
\begin{eqnarray}
G_{mk,lk^{\prime}}&=&\frac{\pi^2 \delta_{l,m \pm 1}}{L^{2}\sqrt {kk^{\prime}}} 
\left[ 
\Lambda^{(1)}_{lm}{\cal P}\frac{k}{k+k^{\prime}}
+i\pi k 
[ \Lambda^{(2)}_{lm} \delta(k-k^{\prime})
-i \pi k \Lambda^{(3)}_{lm} \delta(k+k^{\prime}) ]
-\Lambda^{(4)}_{lm}{\cal P}\frac{k}{k-k^{\prime}} \right] \nonumber \\
&&+\frac{\pi^2 \delta_{l,m}}{L^{2}\sqrt {kk^{\prime}}}
\left[i\frac{\pi}{2}\Lambda^{(2)}_{lm}+(C_i(k-k^{\prime})\Lambda^{(4)}_{lm}
- C_i(k+k^{\prime})\Lambda^{(6)}_{lm} \right]
\label{gij}
\end{eqnarray}
where $C_i(k)$ stands for 
\begin{equation}
C_i(k)=\int\limits_{\lambda}^{L}\frac{\cos kt}{t}dt
\end{equation}
and the set of expressions $\Lambda^{(n)}$ are defined by
\begin{equation}
\Lambda^{(1)}=e^{i\frac{\pi}{2}(l+m)}e^{2i\delta_m}+e^{2i\delta_l}e^{-i\frac{\pi}{2}(l+m)} 
\end{equation}
\begin{equation}
\Lambda^{(2)}=e^{i\frac{\pi}{2}(l - m)}-e^{-i\frac{\pi}{2}(l- m)}e^{2i(\delta_l - \delta_m)}
\end{equation}
\begin{equation}
\Lambda^{(3)}=e^{i\frac{\pi}{2}(l+m)}e^{2i\delta_m}+e^{2i\delta_l}e^{-i\frac{\pi}{2}(l+m)}. 
\label{deff}
\end{equation}
\begin{equation}
\Lambda^{(4)}=e^{i\frac{\pi}{2}(l - m)}+e^{-i\frac{\pi}{2}(l- m)}e^{2i(\delta_l - \delta_m)}
\end{equation}
\begin{equation}
\Lambda^{(5)}=\Lambda^{(2)}+e^{2i\delta_l}e^{-i\frac{\pi}{2}(l+m)}-
e^{-2i\delta_m}e^{i\frac{\pi}{2}(l+m)}
\end{equation}
\begin{equation}
\Lambda^{(6)}=e^{i\frac{\pi}{2}(l+m)}e^{-2i\delta_m}+e^{2i\delta_l}e^{-i\frac{\pi}{2}(l+m)} 
\end{equation}

Expression (\ref{gij}) allows us to write a continuum version of (\ref{scatt}) for the matrix
damping function diagonal elements $S_{i,i}$. Explicitly
\begin{equation}
S_{i,i}(\omega,\omega^{\prime})=\frac{L^4}{4\pi^2}\sum\limits_{ml}
\int k dk \int k^{\prime} dk^{\prime} |G_{mk,lk^{\prime}}|^2 
\delta(\omega - \Omega_k) \delta(\omega^{\prime} - \Omega_{k^{\prime}}),
\label{hhdd}
\end{equation}
where 
\begin{equation}
\omega= \frac{\hbar k^2}{2M}. 
\label{hhd}
\end{equation}

Now, using (\ref{gij})-(\ref{deff}) we can evaluate (\ref{hhdd}) and obtain an 
explicit expression of $S_{i,i}(\omega,\omega^{\prime})$ which, by the time, allow us to 
get the analytic form of $A_{i,i}$ in terms of the variables $\varphi$ and $\eta$ introduced 
before. Then, the limit given by (\ref{chuva}) yields 
\begin{equation}
A_{i,i}(\varphi)=\left(\frac{\pi M}{\hbar}\right)^{2}{\cal G}(\varphi)\varphi^{2} 
\label{agua}
\end{equation}
with 
\begin{equation}
{\cal G}(\varphi)=
2 \sum\limits_{m=1}^{\infty} \sin^2(\delta_{m+1}-\delta_{m}).
\label{trem}
\end{equation}
where the phase shifts $\delta_m$ are also functions of $\varphi$. 

The matrix damping coefficient elements (\ref{land}) can be written now as
\begin{equation}
\overline {\gamma}_{ii}=\left(\frac{8 \pi^2 M}{\hbar}\right)
e^{\beta g \mu_{B} B}
\int\limits_{0 }^{\infty}\frac
{\varphi^{2}e^{\beta\hbar\varphi}{\cal G}(\varphi)}
{(e^{\beta(g\mu _{B}B+\hbar \varphi)}-1)^2}d\varphi.
\label{pvc}
\end{equation}

In computing this final expression (\ref{pvc}) we need a numerical evaluation of the 
phase shift functions defined by (\ref{aene})-(\ref{phase}). This was done taking the 
magnetic field $B=9T$ for three different values of the Land\'e factor $g$ or skyrmion 
size. However, the analysis of the high temperature regime of the skyrmion mobility can 
be done from (\ref{pvc}) without explicit evaluation of ${\cal G}$, giving
\begin{equation}
{\overline \gamma}_{ii}= \frac{M}{\hbar^{3}} 
\left ( \int\limits_{0 }^{\infty} {\cal G}(\varphi) d \varphi \right )
\frac{1}{\beta},
\end{equation}
that increases linearly with temperature independently of the explicit form of 
${\cal G}(\varphi)$.

The behaviour of ${\overline \gamma}_{ii}(T)$ in units of $\gamma_0=\pi^2 \hbar^2 /32 M^2 
\lambda^4$ for any temperature, is shown in Fig.1. for three different values of the 
Land\'e factor. Due to the smallness of the Zeeman energy (0.13 meV, 0.013 meV and 0.0013 meV) 
in all cases the scattering states are excited leading to a damped motion of the skyrmion. 
Although we do not have an analytic expression for ${\overline \gamma}_{ii}(T)$ the numerical 
result shows that near $T=0$ the mobility presents a power law behaviour $T^{\alpha}$, as can 
be seen from the log-log plot in Fig.2, where $\alpha$ is a strong dependent Land\'e factor
coefficient.

\section{conclusions}
Here we have succeded in applying the collective coordinate method of quantum field theory 
to the description of the mobility of the skyrmion excitation in a quantum Hall ferromagnet as 
a function of the temperature. We have been able to treat both elastic and inelastic effects 
of the skyrmion-magnon interaction and show that only the former plays a major role in the 
long time approximation for this problem. We have also shown that in this limit the mobility of the
skyrmion is linear for high temperatures and behaves as $T^{\alpha}$ for low temperatures.

Although we have directly addressed the problem of the quantum Hall ferromagnet, our description 
can be straightforwardly extended to any system which can be mapped into a NL$\sigma$ model or any 
extension thereof. An investigation of the experimental evidences of our findings is now in process.\\

Finally AVF wishes to thank Funda\c{c}\~{a}o de Amparo a Pesquisa do Estado de S\~{a}o Paulo 
(FAPESP) for financial support, whereas AOC kindly acknowledges partial
support from Conselho Nacional de Desenvolvimento Cientifico e Tecnologico (CNPq).

\begin{figure}
\vbox to 9.0cm {\vss\hbox to 9.0cm
 {\hss\
   {\includegraphics{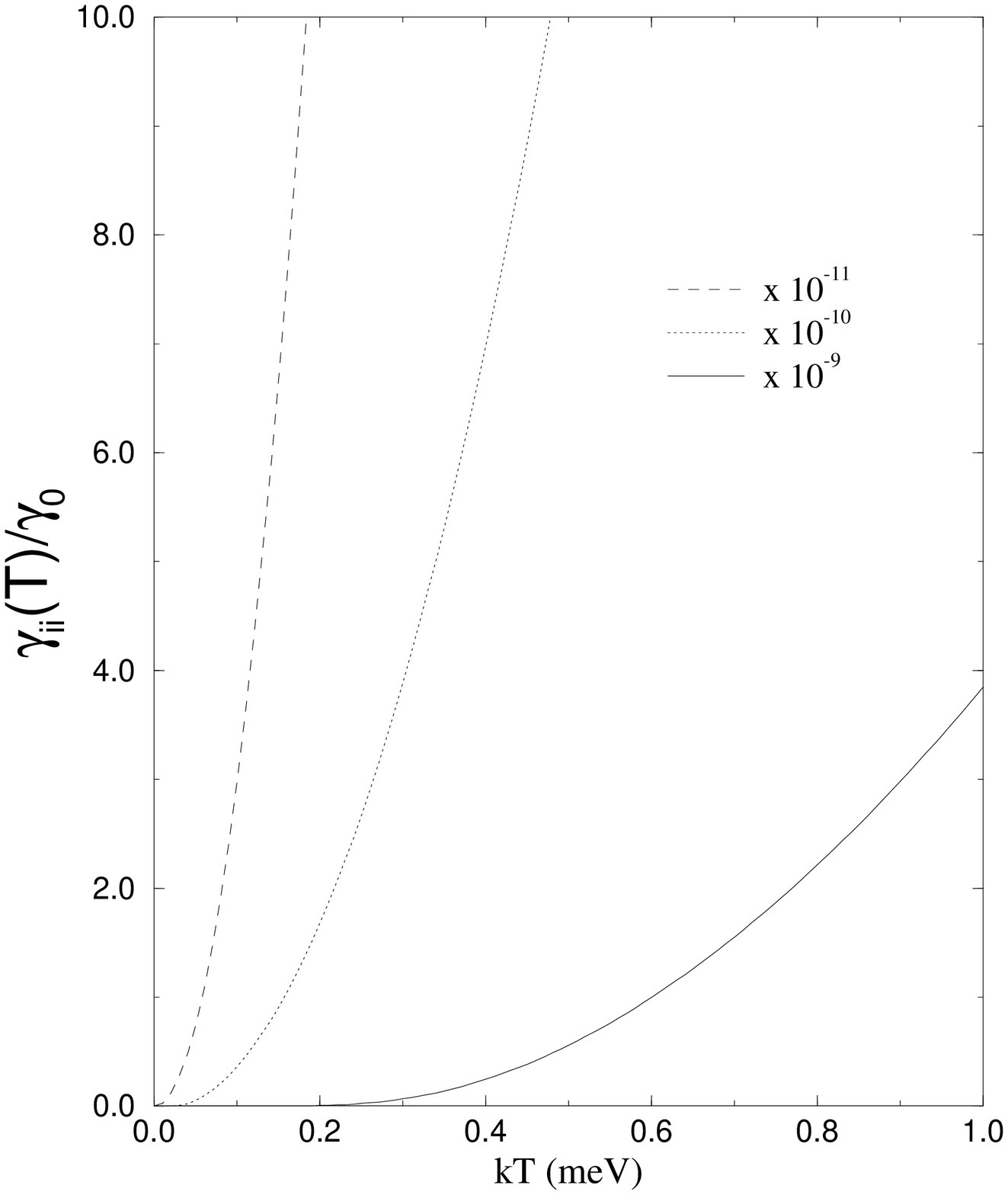}
  }
  \hss}
 }
\caption{Damping coefficient as a function of temperature for different effective Land\'e factors. 
The solid line correspond to the case in which $g=0.25$, the dotted line to $g=0.025$,
and the dashed line to $g=0.0025$}
\label{abs}
\end{figure}

\begin{figure}
\vbox to 9.0cm {\vss\hbox to 9.0cm
 {\hss\
   {\includegraphics{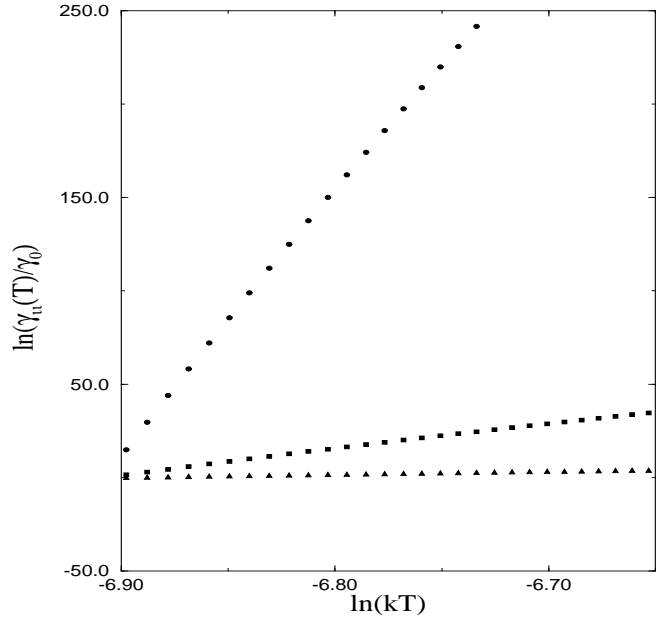}
  }
  \hss}
 }
\caption{ 
The circles correspond to the case in which $g=0.25$, the squares to $g=0.025$,
and the triangles to $g=0.0025$ respectevely.}
\label{abs}
\end{figure}
\end{document}